# Quantum phase modulation with acoustic cavities and quantum dots

Poolad Imany,[1,2,†] Zixuan Wang,[1,2,†] Ryan A. Decrescent,[1] Robert C. Boutelle,[1] Corey A. Mcdonald,[1,2] Travis Autry,[1] Samuel Berweger,[1] Pavel Kabos,[1] Sae Woo Nam,[1] Richard P. Mirin,[1] Kevin L. Silverman[1,*]

[1]National Institute of Standards and Technology, Boulder, CO 80305, USA.
[2]Department of Physics, University of Colorado, Boulder, CO 80309, USA.
*Corresponding author: kevin.silverman@nist.gov.
†These authors contributed equally to this work.

**Abstract**—Fast, efficient, and low power modulation of light at microwave frequencies is crucial for chip-scale classical and quantum processing as well as for long-range networks of superconducting quantum processors. A successful approach to bridge the gap between microwave and optical photons has been to use intermediate platforms such as acoustic waves, which can couple efficiently to a variety of quantum systems. Here, we use gigahertz-frequency focusing surface acoustic wave cavities on GaAs that are piezo-electrically coupled to superconducting circuits and parametrically coupled, via strain, to photons scattered from InAs quantum dots. We demonstrate strong modulation of single photons with a half-wave voltage ($V_\pi$) as low as 44 mV, and subnatural modulation sideband linewidths. These demonstrations pave the way for efficient and low-noise transduction of quantum information between microwave and optical domains.

Electro-optical modulation ranges many classical and quantum applications such as high-speed optical data transmission [1,2] and quantum transduction of information between light and matter qubits [3,4]. An outstanding challenge for microwave-to-optical conversion of photons arises from the energy (frequency) difference between these two domains, exceeding five orders of magnitude. To link microwaves and optics efficiently, the use of acoustic phonons as an intermediate platform has been proposed [3,5–9]. The slower speed of sound compared to light (~ five orders of magnitude) results in smaller wavelength of phonons than that of photons for the same frequency. For this reason, phonons can couple more efficiently to a variety of quantum systems, such as superconducting circuits [3,10], defect centers in diamond [11], and semiconductor quantum dots (QDs) [12,13], and are emerging as universal transducers [14].

Here, we use acoustic cavities as an intermediate platform for electro-optical interactions and use quantum emitters in the form of QDs. The use of QDs ensures that only single photons are emitted in a transduction cycle. We choose focusing surface acoustic wave (SAW) cavities on GaAs, which can host InAs QDs. SAWs on GaAs are a favorable platform in that they couple to superconducting microwave circuits piezo-electrically, and to InAs QDs via strain. This platform supports critical electro-acoustic coupling and high-quality-factor cavities with small mode volumes, leading to significant single phonon coupling rates ($g_0$), and narrow scattering linewidths well below the lifetime limit of the QD [15]. The development of state-of-the-art SAW cavities and the characteristics of our chosen platform facilitate strong electro-acousto-optic modulation, yielding a half-wave voltage ($V_\pi$) of 44 mV, a total photon conversion efficiency of $\eta = 10^{-15}$, and a $g_0$ of $2\pi \times 16$ kHz with a clear path towards megahertz rates.

Figs. 1(a,b) show an illustration of our device and its scanning electron microscope (SEM) image, respectively, and the experimental setup is depicted in Fig. 1(c). Acoustic distributed Bragg reflectors etched into the surface of GaAs form the focusing SAW cavity, driven by a set of niobium interdigitated transducers (IDTs) inside the cavity. Both structures are designed for microwave frequencies around 3.6 GHz and are fabricated on a wafer with epitaxially grown InAs QDs with characteristic photon emission rate of $2\pi \times 200$ MHz. We perform a reflection measurement of an acoustic mode at 4 K in Fig. 2(a), showing a reflection dip of ~ 9 dB on resonance. Fits to the measured microwave reflection yield an internal (external) quality factor of $Q_i = 7.6 \pm 0.1$ k ($Q_e = 15.3 \pm 0.3$ k), demonstrating near critical coupling.

For acousto-optic characterization of the device, we shine tightly focused non-resonant light ($\lambda = 632$ nm) at different positions inside the SAW cavity, and measure the quantum dot ensemble emission using a spectrometer with and without a microwave drive at the SAW cavity mode frequency. As the SAWs modulate the QD transitions, the QD spectrum spreads and the sharp transition lines broaden (Supplementary Information). We develop a simple metric to quantify this spreading at each position and map the SAW cavity mode using QDs as local strain gauges [Fig. 2(b), Supplementary Information]. The acoustic waist measured for the device shown in Fig. 1(b) is 4.6 μm, in agreement with the designed value (Supplementary Information). In Fig. 2(b), we measure a different device with a narrower SAW waist, where scans at different x positions along the cavity length confirm the focusing profile of the SAW cavity mode. We then focus our attention on individual QDs at the acoustic waist of the cavity, where the acousto-optic interaction is the strongest. A single QD transition is isolated and pumped resonantly ($\lambda \sim 912$ nm), while the SAW cavity is driven at the cavity mode frequency with different microwave powers. Phonons modulate the photons scattered from the QD, shifting their frequency by increments of the SAW frequency [12]. We use a cross-polarization setup to reject the scattered pump photons by at least six orders of magnitude [16], and use a tunable Fabry-Perot filter (600 MHz linewidth) and a superconducting nanowire single-photon detector (SNSPD) to measure the spectrum of the scattered photons [Fig. 3(a)]. The single photons detected to the red (blue) side of the resonant laser herald phonons being added to (subtracted from) the SAW cavity. The spectra were fit to obtain the microwave power-dependent modulation index, δ. We extract a $V_\pi$ of 44 mV, comparable to the state-of-the-art modulators [5,7,8,17]. This low $V_\pi$ is a result of the strong electro-acoustic coupling afforded by the SAW cavity, followed by efficient photon-phonon coupling mediated by the QD. To demonstrate the SAW cavity-enhanced aspect of the interactions, we repeat the experiment while driving SAWs at a frequency 2 MHz away from the cavity mode, which leads to a $V_\pi$ of 220 mV, five times higher than that for the cavity-enhanced process [Fig. 3(b)]. The lower $V_\pi$ when driving the cavity mode is due to improved impedance matching of the IDT, as well as enhanced acousto-optic coupling due to the finesse of the cavity ($\mathcal{F} = 14$). For this device, we also calculate the single phonon coupling rate of $g_0 = 2\pi \times 1.4$ kHz. This modest value results from the QDs being 750 nm below the surface where the strain field of the SAWs is relatively small (Supplementary Information, [18]).

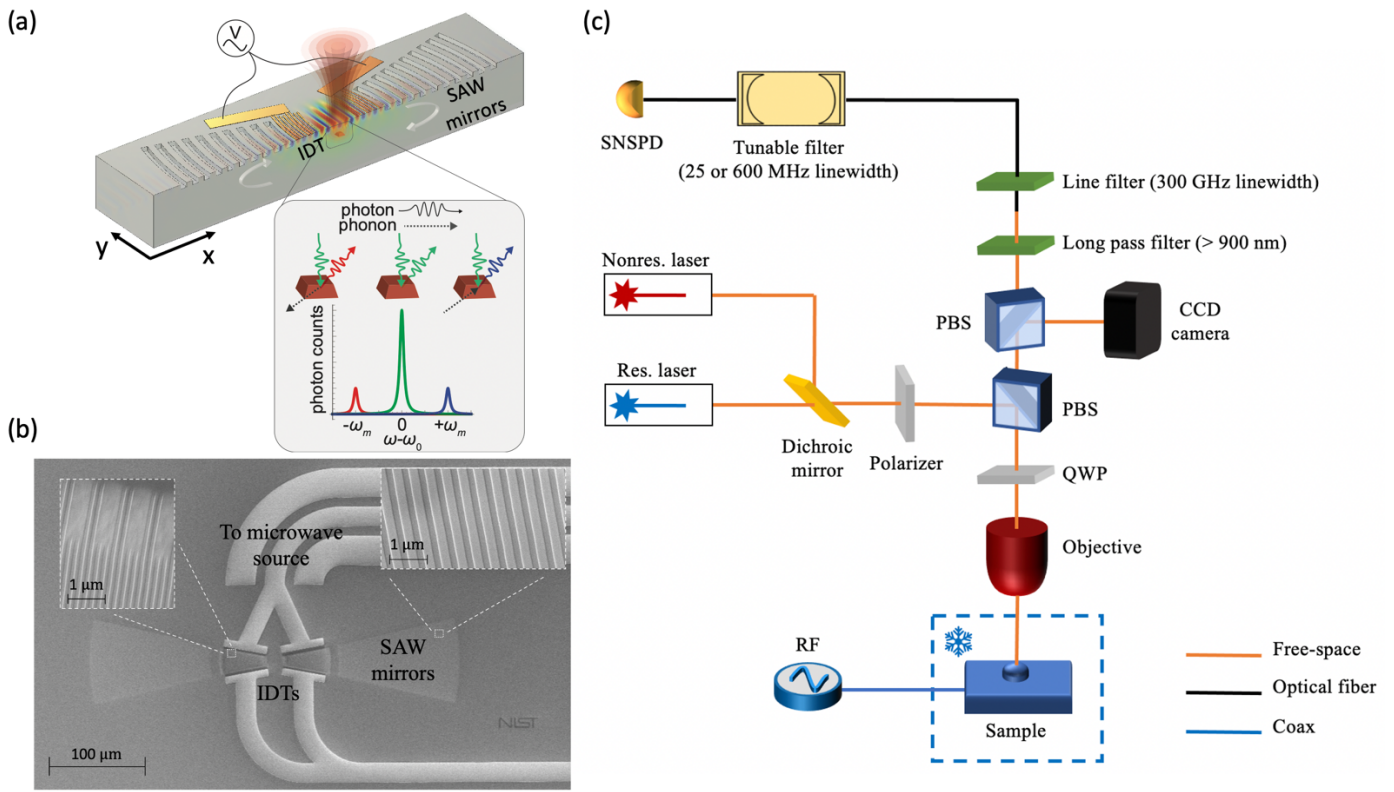

**Fig. 1.** (a) Illustration of the modulation scheme. An IDT drives a SAW cavity to generate phonons from microwave photons. The phonons interact with optical photons mediated by a QD, generating sidebands in the scattering spectrum. The inset shows three different photon scattering processes: (left) one phonon is emitted; (middle) no phonon is involved; (right) one phonon is absorbed. The x axis for the inset is centered at the QD transition. (b) Scanning electron microscopy image of the fabricated device. (c) The experimental setup. PBS: polarizing beam splitter. QWP: quarter waveplate. RF: radio frequency. CCD: charge-coupled device. SNSPD: superconducting nanowire single photon detector.

Next, we show that, as required for quantum transduction, the coherence properties of scattered photons are likely determined by that of the acoustic mode and the pump laser, and not by incoherent emission from the QD. Resonant scattering from QDs for low pump powers follows the linewidth of the laser, not that of the QD [15]. Indeed, using a much narrower tunable filter (25 MHz linewidth), we observe sideband linewidths that are equal to the filter [Fig. 3(c)], confirming that our sidebands are much narrower than the natural lifetime-imposed limit of our two-level system. Narrow-linewidth sidebands indicate that they can be filtered rigorously for low-noise quantum transduction purposes. Additionally, due to the multimode nature of our SAW cavity, a coherent superposition of two or more cavity modes can be used and filtered simultaneously, giving our device the ability to transduce frequency qubits [19]. We also note that our QDs are single photon emitters, and our device is a source of modulated *single photons* [20] that can be readily used for quantum communications protocols.

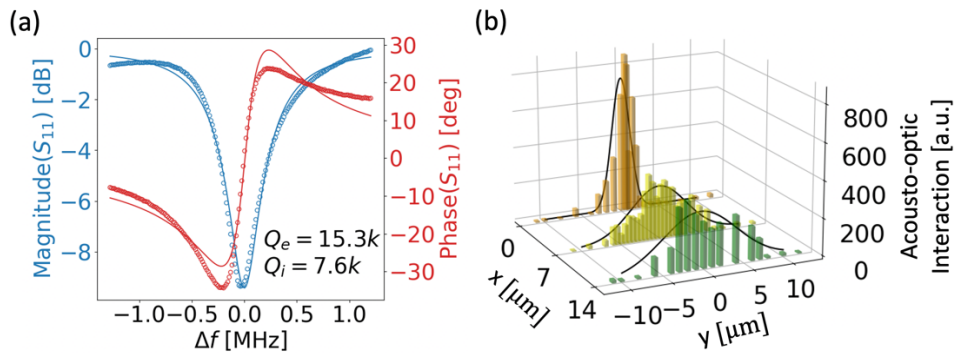

**Fig. 2.** (a) Microwave reflection spectrum of the SAW cavity mode (open circles). The solid curves are fits to the data. (b) Optical measurement of the acoustic waist (along the y axis) inside a focusing cavity at different distances along the cavity length (x axis, Supplementary Information). The beam center is positioned at (x,y)=(0,0). The black curves are Gaussian fits to the data at each y value. At the focus, the measured SAW waist is 2.4 μm.

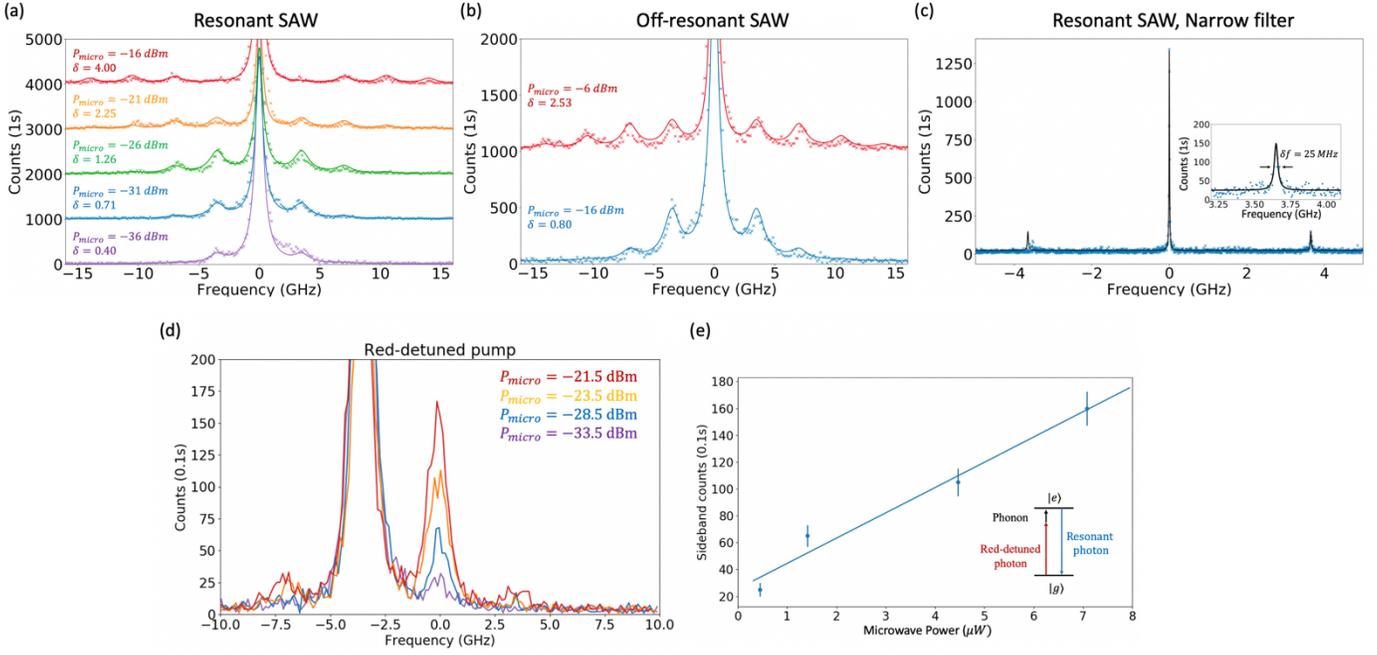

**Fig. 3.** (a),(b) Modulation of resonantly scattered light when driving the SAW cavity (a) on resonance and (b) off resonance, with varying microwave powers. The x axis is centered at the QD transition frequency. The center peaks are mostly due to residual pump photons. $P_{micro}$ is the microwave input power driving the IDT, and δ is the modulation index. The data for different $P_{micro}$ are offset in the vertical direction for clarity. The solid curves are fits to the data, considering that the modulation index scales as $\sqrt{P_{micro}}$. The microwave frequency in (b) is 2 MHz off resonance from the cavity mode. The slight asymmetry in the modulation spectra is due to mismatch between the pump light and QD transition frequency, similar to optical cavities [23]. (c) The center and first two sideband peaks scanned with a narrower filter (25 MHz linewidth), showing sub-natural sideband linewidths. The asymmetry in the position and counts of the sidebands is due to temperature fluctuations of the filter over the course of the measurement. (d) Optical pump red-detuned by one SAW frequency. The large peak at -3.6 GHz is mostly due to residual pump photons. We see strong asymmetry in the sidebands, indicating that the photon scattering process on average removes phonons from the SAW cavity. (e) Resonant scattering counts for red-detuned optical pumping, as a function of input microwave power.

To show single-phonon transduction, we red-detune the laser by one SAW frequency and tune the microwave power such that the dominant process is the absorption of a single phonon and thus emission of a resonant photon [Fig. 3(d)]. We observe that the resonant photon counts scale linearly with the SAW power [Fig. 3(e)], showing only one phonon is involved in this low-power scattering process. From Fig. 3(e), we can calculate the total efficiency of detecting an optical photon upon driving the SAW cavity with a single microwave photon, yielding $\eta = 10^{-15}$. We can enhance $\eta$ in the future by creating open photonic structures around the QD, growing the QDs closer to the surface, controlling the charge state of the QD, and fabricating SAW cavities with smaller mode volumes (Supplementary Information).

In conclusion, we have integrated superconducting electrical circuits, surface acoustic wave resonators and quantum emitters into a single device to generate efficient electro-optic interactions, and observed strong gigahertz-regime modulation of single photons. These preliminary results open the door for demonstrations such as quantum transduction, sideband cooling [6,21] of an acoustic cavity mediated by a quantum emitter, and photon-phonon entanglement generation [22].

**Funding.** National Institute of Standards and Technology (NIST), National Research Council (NRC).

**Acknowledgments.** The authors thank John Teufel and Konrad Lehnert for fruitful discussions, and Varun Verma for fabrication of the superconducting nanowire single photon detector.

# Supplementary Information

1. **Device fabrication**

The quantum dot (QD) sample is grown by molecular beam epitaxy (MBE). 1.65 monolayers of InAs QDs are sandwiched in a distributed-Bragg-reflector (DBR) optical cavity with 12 (below the QDs) and 4 (above the QDs) pairs of alternating - $Al_{0.95}Ga_{0.05}As$/GaAs layers, as shown in Fig. S1(a). The MBE-grown wafer is then processed for fabrication of surface acoustic wave (SAW) cavities and interdigitated transducers (IDTs). An $SiO_2$ hard mask is sputtered on the wafer to allow multiple etch depths on a single wafer and to reduce re-sputtering contamination of the resist during the etching process. The SAW cavity trenches are defined by electron beam lithography (EBL). The exposed pattern is developed in 1:3 MIBK/IPA solution at 5 °C. Using reactive-ion etching (RIE), the SAW cavity pattern is etched first into the hard mask using $CHF_3$, then into GaAs using $BCl_3$. Subsequently, the hard mask is stripped with hydrofluoric acid, and a 20 nm layer of Nb is deposited on the surface. The IDT is fabricated in two primary steps. First, the negative image of the IDT fingers is defined by EBL and etched away with RIE ($SF_6$). Second, the negative image of the IDT trace (including the wirebond pad and the coplanar waveguide) is defined by photolithography and etched away also with RIE ($SF_6$). The IDT fingers are covered with photoresist during the second etch step. The main steps are summarized and illustrated in Fig. S1(b). Since niobium has low conductivity at room temperature, a 100 nm layer of aluminum is deposited on the IDT trace using a liftoff process to allow measurements at room temperature as well as cryogenic measurements.

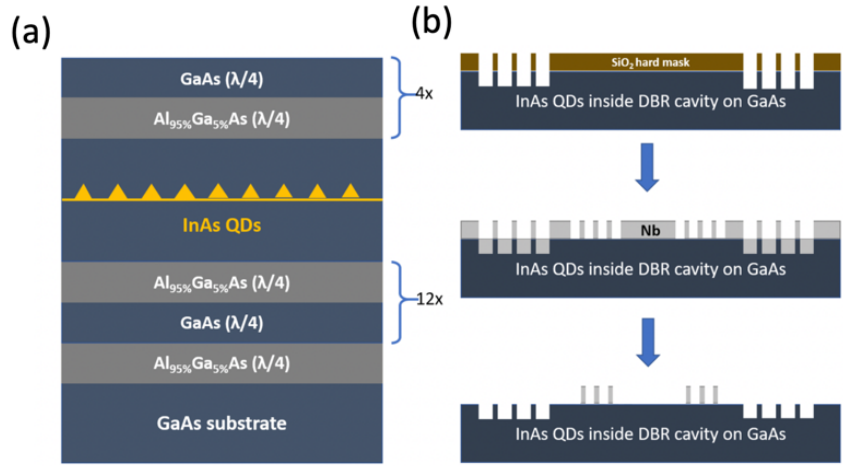

**Fig. S1.** (a) The structure of the sample grown by molecular beam epitaxy (MBE). (b) Fabrication steps after the MBE growth.

2. **Experimental details**

A simplified version of the experimental setup is depicted in Fig. 1(c). On the optical side, we use a polarization-based dark-field confocal microscopy technique, following closely the design from ref [1]. A resonant tunable laser (Toptica CTL, 910-980 nm) is combined with a non-resonant pump (HeNe) on a dichroic mirror. The polarization is then cleaned up to ~8 orders of magnitude with a linear polarizer (Thorlabs). The beam is then reflected from a polarizing beam splitter (PBS), focused by an objective with NA=0.7 (Mitutoyo 100x), and enters a cryostat (Montana Instruments) held at 4K where it illuminates the sample. The SAW cavity is driven electrically using an IDT, with either a vector network analyzer (VNA) (Copper Mountain) to locate the cavity modes, or a multimode RF synthesizer (Holzworth) that can produce sinusoidal signals with +23 dBm of power up to 5 GHz. The light reflected from the sample goes through a zero-order quarter waveplate to compensate for polarization aberrations, and two PBSs rejecting the resonant pump laser intensity by more than six orders of magnitude. A CCD camera is mounted at the reflection port of the second PBS to image the sample and to position the focused pump beam at the center of the SAW cavity. A long pass filter rejects the non-resonant pump. The collected light is fiber coupled, then recollimated to pass through a tunable line filter (Alluxa) to isolate the transition of a single QD. Finally the light goes through one of the two choices of Fabry-Perot etalon tunable filters with wide (600 MHz) or narrow (25 MHz) linewidths (LightMachinery and Thorlabs, respectively), and is detected by a superconducting nanowire single-photon detector (SNSPD).

## 3. Experimental and theoretical estimates of the single-phonon coupling rate, $g_0$

### 3a) Relating the single-phonon coupling rate to the classical modulation index

The modulation index, $\delta$, is derived by studying the resonance fluorescence spectrum of single QDs driven by microwave-frequency SAWs. The power spectrum of QD resonance fluorescence is given by [2]:

$$P[\omega] = \sum_{n=-\infty}^{\infty} \frac{J_n^2(\delta)}{\gamma^2 + [\omega - (\omega_0 - n\omega_m)]^2} \quad (S1)$$

where $J_n(\delta)$ is the $n$-th order Bessel function, $2\gamma$ is the full width at half maximum of the Lorentzian QD emission, $\omega_0$ is the QD's center frequency, and $\omega_m$ is the SAW cavity driving frequency (and thus the frequency of the QD's energy modulation). Experimental resonance fluorescence spectra are fit to obtain $\delta$.

We would like to relate the experimentally obtained values for $\delta$, which depend on input microwave powers [e.g., see Fig. 3(a) in the main text], to intrinsic single phonon-QD interaction rates. Physically, $\delta$ describes the magnitude of the QD's energy shift under modulation at frequency $\omega_m$ according to [2]

$$\omega_{QD} = \omega_{QD,0} + \delta\omega_m \sin(\omega_m t) \quad (S2)$$

In a quantum-mechanical treatment, the interaction arises from a Taylor-series expansion of the QD energy to first order in the strain (or displacement) amplitude according to:

$$\omega_{QD} = \omega_{QD,0} + \left.\frac{\partial \omega_{QD}}{\partial \epsilon}\right|_{\epsilon=0} \epsilon \equiv \omega_{QD,0} + g_0(u/u_{zpm}) \quad (S3)$$

where the strain, $\epsilon$, and the displacement, $u$, are related by $\epsilon = iku$ for a mechanical wave with momentum $k$ in a scalar treatment. The strain and displacement amplitudes are to be treated as operators; specifically, $\hat{u}$ is the displacement (position) operator of the mechanical mode in question and can be written in terms of creation and annihilation operators ($\hat{b}^\dagger$ and $\hat{b}$) and the zero-point amplitude ($u_{zpm}$): $\hat{u}=u_{zpm}(\hat{b} + \hat{b}^\dagger)$. The factor $g_0$ thus describes the QD-phonon interaction rate scaled to the single-phonon level. The operators $\hat{b}$ and $\hat{b}^\dagger$ return eigenvalues $\sqrt{n}$ and $\sqrt{n+1}$ when acting on phonon number states, respectively. For small steady-state phonon numbers, the blue and red sidebands observed in resonance fluorescence therefore have unequal weighting. However, for a large steady-state phonon population, $n \gg 1$, such that $\sqrt{n} \simeq \sqrt{n+1}$, the sideband weights become approximately equal and a direct comparison of Eqns. S2 and S3, assuming a sinusoidal time dependence of the strain field in Eqn. S3, provides the relationship [3]:

$$\delta = 2g_0\sqrt{n}/\omega_m \quad (S4)$$

The microwave driving frequency $\omega_m$ is chosen to excite the desired SAW cavity mode. The steady-state cavity phonon number, $n$, is determined from the microwave driving power, $P_{micro}$, and the total (external) cavity coupling rates $\kappa(\kappa_{ext})$, through the relationship:

$$n = \frac{P_{micro}}{\hbar\omega_m} \frac{4\kappa_{ext}}{\kappa^2} \quad (S5)$$

The coupling rates are calculated from the microwave reflection ($S_{11}$) spectra. $V_\pi$ is the microwave voltage used to drive the SAW cavity, inferred from the microwave power using $V = \sqrt{2ZP_{micro}}$ and $Z = 50\Omega$, such that $\delta = \pi$. $V_\pi$ is an important metric since it shows the voltage required for the scattered photons to be out of phase with the incident light. We measure a 3.5 dB loss between our RF generator and the sample mount inside the cryostat, which is calibrated out when reporting $P_{micro}$ in Fig.3 and Fig. S4.

### 3b) Theoretical estimates of $g_0$

From Eqn. S3, the single-phonon coupling rate $g_0$ is written $g_0 = ik(\partial\omega_{QD}/\partial\epsilon)|_{\epsilon=0} u_{zpm}$. The factor $G = (\partial\omega_{QD}/\partial\epsilon)|_{\epsilon=0}$ is inherent to the material system in question (the GaAs/InGaAs QDs), and $k$ depends on the wavelength of the SAW mode. Importantly, $u_{zpm}$ is the zero-point amplitude of the SAW mode *at the position of the QD*, and its scale depends on SAW cavity mode volume. We calculate $u_{zpm}$ by equating the classical energy of the SAW cavity mode to the quantum zero-point energy of the mode [4]. We write the zero-point displacement field, $\vec{u}_{zpm}(x,y,z)$, as:

$$\vec{u}_{zpm}(x,y,z) = u_{0,\text{zpm}}\{u_x(x,y,z), u_y(x,y,z), u_z(x,y,z)\} \quad (S6)$$

where $u_{0,\text{zpm}} = \sqrt{\hbar/2\rho\omega V}$ and $V = \int d^3x\, |\{u_x(x,y,z), u_y(x,y,z), u_z(x,y,z)\}|^2$. For a paraxial Gaussian standing-wave SAW mode confined along the $x$ direction and focusing in the $y$ direction at a $z$-normal GaAs surface, we approximate the normalized field components as:

$$u_x(x,y,z) = -\frac{1}{3\sqrt{2}\,\kappa} e^{ik\kappa z} \frac{w_0}{w(x)} e^{-\frac{y^2}{w(x)^2}+ik\frac{y^2}{2R(x)}-\tan^{-1}\left(\frac{x}{x_R}\right)} (e^{ikx} - e^{-ikx})$$

$$u_y(x,y,z) = 0$$

$$u_z(x,y,z) = e^{ik\kappa z} \frac{w_0}{w(x)} e^{-\frac{y^2}{w(x)^2}+ik\frac{y^2}{2R(x)}-\tan^{-1}\left(\frac{x}{x_R}\right)} (e^{ikx} + e^{-ikx}) \quad (S7)$$

The Gaussian beam parameters are defined in the usual manner: $x_R = \pi w_0^2/\lambda$, $w(x) = w_0\sqrt{1+(x/x_R)^2}$, and $R(x) = x[1+(x_R/x)^2]$. For SAWs at ~3.6 GHz along the GaAs (110) direction, we take $k = 2\pi/\lambda$ with $\lambda = 0.830$ μm and $\kappa = 0.483 - i0.489$. We calculate $\vec{u}_{zpm}(x,y,z)$ for each device independently by inputting the measured $w_0$ and numerically evaluating the volume integral over a range in $x$ constrained to the measured cavity length. We assume a single (scalar) theoretical value for the QD deformation potential from ref. [2]: $G = 6.5 \times 10^{14}$ Hz. Fig. S2 shows the resulting $x$-dependent values for $g_0$, evaluated at the known depth ($z$) of the QD and along the beam center ($y=0$). Specifically, we plot $f(x) = |\text{Re}[ikGu_{0,\text{zpm}}u_x(x,0,z_{QD})]| + |\text{Re}[ikGu_{0,\text{zpm}}u_z(x,0,z_{QD})]|$ near the beam focus ($x=0$) over a single SAW wavelength (solid black curve). Blue and orange curves show the individual $u_x$ and $u_z$ contributions. Experimentally, the exact position of the QD within the standing-wave pattern is not known, and the $y$-position may be on the order of 1μm away from the beam center.

4. **SAW cavity coupling enhancement**

The SAW cavity enhances both acousto-optic coupling (between the SAWs and the QD) and electro-acoustic coupling (between the IDT and the SAWs), since both the QD and IDT are inside the cavity. The overall enhancement is calculated by comparing the results presented in Figs. 3(a, b) of the main text, which a factor of five reduction in $V_\pi$, and consequently, a factor of 25 reduction in the required microwave power to obtain the same modulation spectrum, for the cavity-enhanced interaction, compared to the same system driven with off-resonance microwaves. The acousto-optic contribution of this enhancement is due to multiple internal reflections of resonant SAWs inside the cavity. This can be quantified by finesse, which is 14 for the device presented in Figs. 3(a,b). The electro-acoustic coupling enhancement is due to improved impedance matching to the IDT around the SAW cavity resonance, which is responsible for the remaining microwave power reduction of 25 compared to the acousto-optic enhancement of 14.

5. **Different device architectures**

We fabricated four different devices and characterized them electro-mechanically, opto-mechanically, and electro-optically. These devices vary in a range of parameters such as IDT finger number, SAW cavity mode volume and waist, and the presence of a distributed Bragg reflector layer on top of the QD layer forming an optical cavity. In this section, we show the design for each of these devices, along with a detailed analysis of these parameters and finally, summarize them in Table S1. The designs are shown in Fig. S3.

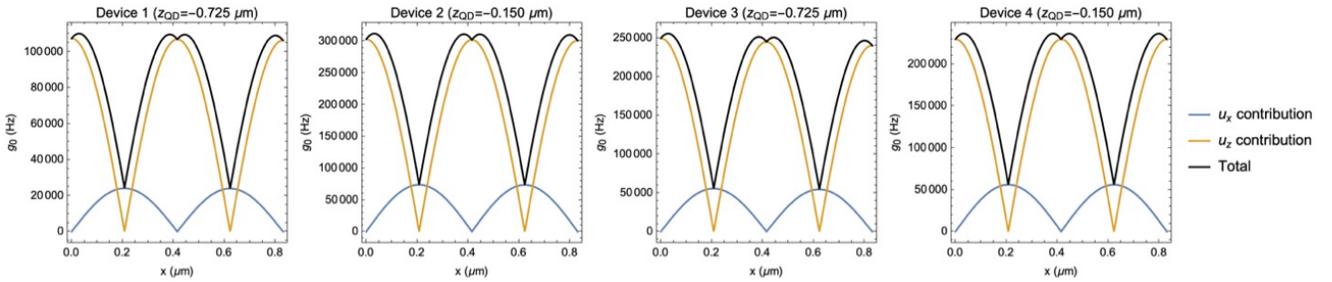

**Fig. S2.** Calculated $g_0$ values along the standing-wave pattern near the center of the cavity for four different device architectures reported in section 5.

- **Dev1**

This device was fabricated on a wafer with distributed Bragg reflectors (DBRs) both at the bottom and top of the QD layer (Wafer1). Dev3 and Dev4 are also from Wafer1, with four pairs of DBR on top of the QD layer, resulting in the QDs being buried 500 nm below the surface, which is comparable to the wavelength of the SAWs (820 nm), and will reduce the strain by a factor of 10 compared to the surface. However, the optical cavity formed by the DBRs around the QDs result in more efficient resonant pumping and collection of the scattered light. The QDs on Wafer1 are rather dense, such that at every focusing spot on the sample, at least 20 QD lines are observable on the spectrometer. Many of these lines however, could belong to different charge states of the same QDs.

Dev1 has a tightly focusing SAW cavity design, with a SAW waist of 4.6 µm at the center and an effective cavity length of 146 µm. There are two IDTs on both sides of the cavity, each with 50 periods, driven with the same microwave, same as in Dev3. This is the device we used in Fig. 3 of the main text. SAW cavity resonances in the $S_{11}$ spectra were observed with dips as large as 9 dB. We measure a $g_0$ of 1.4 kHz, which is lower than our estimate of $g_{est,max} = 110$ kHz, calculated in Sec. 3(b). The discrepancy between the measured $g_0$ and calculated $g_{est,max}$ is still under investigation.

- **Dev2**

This device was fabricated on a different wafer (Wafer2). The quantum dots are closer to the surface (150 nm) which results in a larger $g_0$ because of the larger strain amplitude there. The effective cavity length for this device is 229 µm, and the SAW waist is 5.6 µm. However, the QDs on this dot were more sparse, and the only region with QDs were discovered in the cavity is 100 µm away from the focus where the SAW waist is 15 µm. The IDT for Dev2 is only on the right side of the cavity and has 100 periods. The modulation spectrum for this device is shown in Fig. S4(a). Cavity resonances dips were approximately 0.3 dB.

From Wafer2, resonance fluorescence is not observed without the help of a small amount of non-resonant light, which is well known to improve the charge state of the QD. However, this results in some undesired background photon counts [Fig S4(a), orange curve]. For this device, we measure a $g_0$ of 16 kHz which is higher than that of Dev1 because the QDs are closer to the surface (150 nm vs 750 nm deep) ($g_{est,max} = 310$ kHz).

- **Dev3**

Fabricated on Wafer 1, this device has two 25-period IDT and an effective cavity length of 123 µm. This cavity is more tightly focusing and the SAW waist is 2.4 µm, resulting in a higher $g_0$ compared to Dev1 by about a factor of 2. The modulation spectrum for this device is shown in Fig. S4(b). Cavity resonance dips were approximately 0.1 dB. We suspect the small cavity dip of this device is largely due to long wirebonds leading to poor impedance matching. We measure a $g_0$ of 3.2 kHz ($g_{est,max} = 260$ kHz), which is more than a factor of 2 higher than that of Dev1, due to tighter focusing of the acoustic energy.

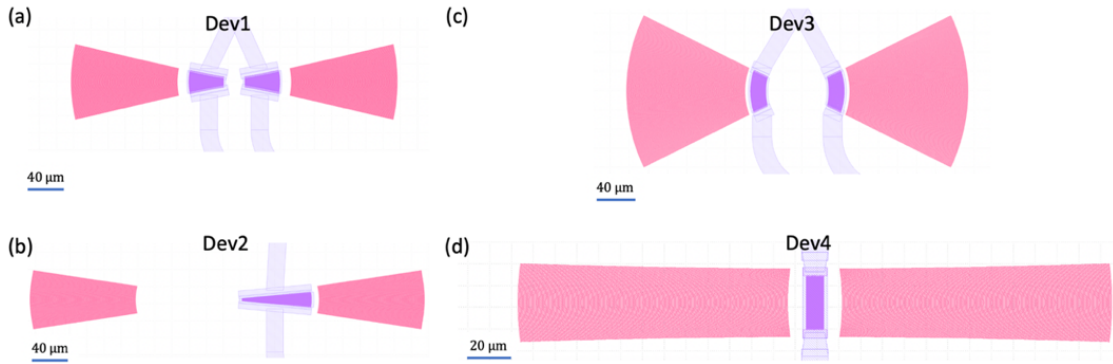

**Fig. S3.** The SAW cavity and IDT designs for four devices used in experiments. Pink indicates mirror regions. Purple indicates IDTs and electrical connections.

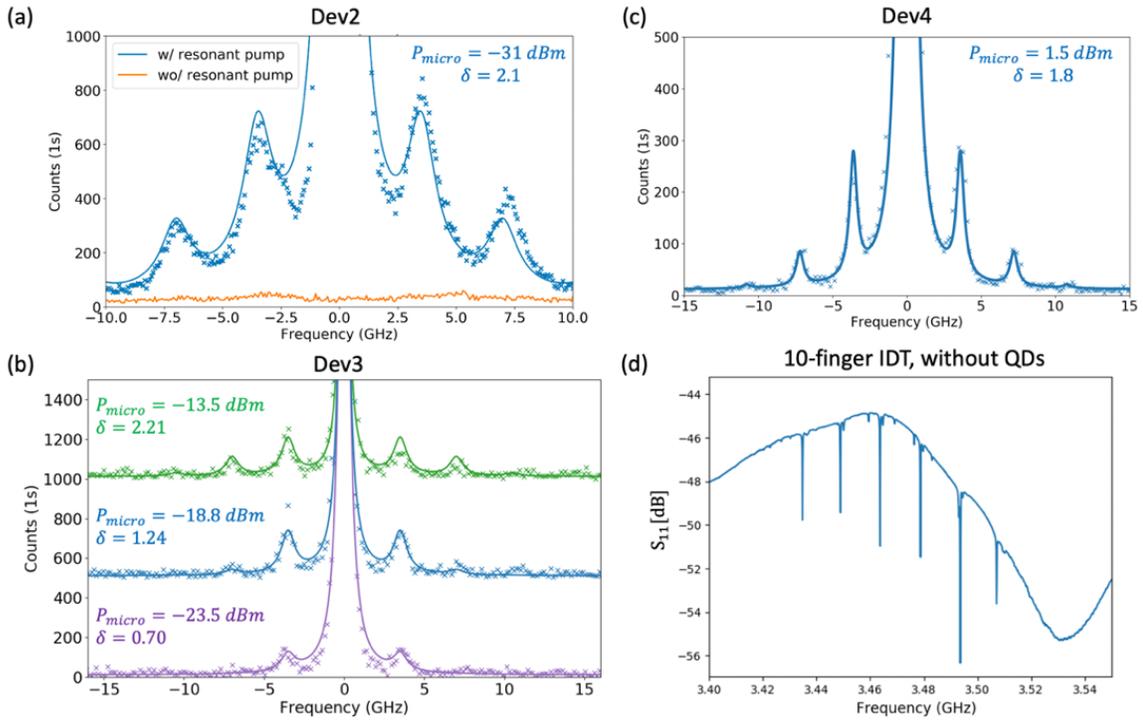

**Fig. S4.** (a-c). Modulation spectrum for resonant pumping of the quantum dots for different devices. (d) The $S_{11}$ measurement for a SAW cavity with a 10-period IDT inside, fabricated on a dummy wafer without QDs, showing strong electro-acoustic coupling.

- **Dev4**

This device has a weakly focusing SAW cavity (beam waist of 25 µm) and a planar IDT with only 10 periods at the center. The small number of IDT fingers allows us to fabricate the SAW cavity with a smaller cavity length (39 µm), which results in a slightly larger mode volume and smaller $g_0$ compared to Dev1, despite its large waist. The modulation spectrum of Dev4 is shown in Fig. S4(c). These results and comparisons between the four devices are summarized in Table S1.

Due to very small number of IDT fingers and roughness of the GaAs surface for this device, this device has a very small cavity dip in the $S_{11}$ measurement (0.01 dB). For the next generation of devices, we have upgraded to an IDT fabrication recipe which leaves less rough surfaces around the IDT, and has resulted in cavity dips of more than 8 dB for even 10-finger IDTs [Fig. S4(d)]. However, these devices were fabricated on a wafer without QDs. We expect to fabricate SAW cavities with small mode volumes while maintaining a good electro-acoustic coupling in near future.

Table S1. Comparison between different devices.

| Device | Pros | Cons | $g_0/2\pi$ (kHz) | $g_{est,max}/2\pi$ (kHz) |
|---|---|---|---|---|
| Dev1 | • Best electro-mechanical coupling |  | 1.4 | 110 |
| Dev2 | • Best SAW-QD interaction | • Poor photon collection, <br> • Poor electro-mechanical coupling | 16 | 310 |
| Dev3 | • Narrow SAW waist | • Poor electro-mechanical coupling | 3.2 | 260 |
| Dev4 | • Short cavity length | • Wide SAW waist, <br> • Poor electro-mechanical coupling | 1.4 | 240 |

## 6. Quantifying surface acoustic wave beam profiles by quantum-dot ensemble photoluminescence smearing

Since changes to the QD fluorescence spectrum depend on the local strain amplitude, we can quantify the strain profiles of our SAW cavities through changes in the fluorescence spectra at a series of positions within the cavities. Rather than studying single QD resonance fluorescence, we study the ensemble photoluminescence (PL) of non-resonantly pumped QDs with and without SAW driving. Example spectra are shown in Figs. S5(a,b); blue curves correspond to PL spectra when driving the SAW cavity with 20 dBm of microwave power ('SAWs on') while gray curves correspond to reference PL spectra from the same position without microwave driving ('SAWs off'). Individual QD emission lines are observable, and we estimate that tens of QDs contribute to each spectrum originating from the diffraction-limited pump spot. When far from the cavity center [~10 μm, Fig. S5(a)], spectra with 'SAWs on' are virtually identical to reference spectra with 'SAWs off', indicating no QD modulation. At the center of the cavity [Fig. S5(b)], there are noticeable changes to the spectra when driving SAWs.

We quantify the local modulation strength by defining a metric that quantifies the difference between the 'SAWs off' and 'SAWs on' spectra. Our chosen metric is motivated by the Kolmogorov-Smirnov (KS) test where our reference function is taken to be the measured 'SAWs off' reference spectrum. Specifically, at each position, we first define a normalized cumulative distribution function, $C(\lambda)$, from each PL spectrum, $PL(\lambda)$:

$$C_{\text{off}}(\lambda) = \int_{-\infty}^{\lambda} PL_{\text{off}}(\lambda')d\lambda' \Big/ \int_{-\infty}^{\infty} PL_{\text{off}}(\lambda')d\lambda' \quad (S8)$$

$$C_{\text{on}}(\lambda) = \int_{-\infty}^{\lambda} PL_{\text{on}}(\lambda')d\lambda' \Big/ \int_{-\infty}^{\infty} PL_{\text{on}}(\lambda')d\lambda' \quad (S9)$$

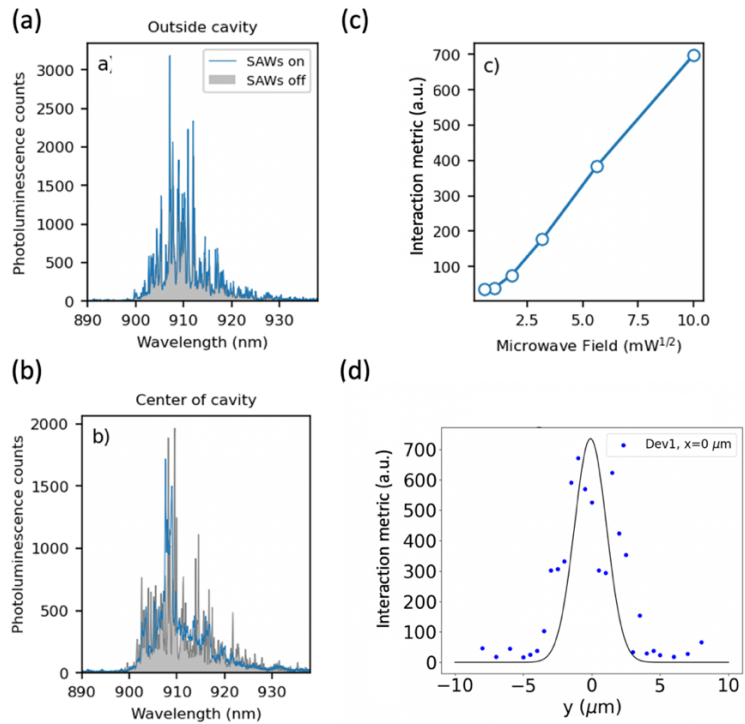

**Figure S5.** PL detection of the local SAW strain amplitude. (a) PL spectra from an ensemble of QDs approximately 10 μm (approximately 4 beam waists) from the SAW cavity focal point. Gray filled curve is the "reference" PL spectrum recorded without driving the SAW cavity with microwaves; blue curve is the PL spectrum recorded from the same position while driving the SAW cavity at 20 dBm microwave power. (b) Same as (a), but for a position very close to the SAW focal point (center of cavity). (c) The interaction metric D for spectra recorded near the center of the cavity for various driving microwave field amplitudes. (d) Resulting SAW strain profile at the center of the SAW cavity for Dev1. The black curve is the Gaussian fit to the data.

We then quantify the difference between the two spectra by integrating the absolute value of the difference of the cumulative distribution functions:

$$D = \left[\int_{-\infty}^{\infty} |C_{\text{off}}(\lambda) - C_{\text{on}}(\lambda)| d\lambda\right]^2 \quad (S10)$$

This specific form of $D$ was determined through a calibration test by measuring $D$ at a single position over a controlled range of microwave driving powers. We find that $D$ as defined is approximately proportional to the microwave driving field and thus the local strain [Fig. S5(c)]. We thus measure the strain profiles of our SAW cavities by measuring two spectra, and calculating $D$, each at a series of points spanning the SAW cavity mode.

This metric was used for Fig. 2(b) of the main text for Dev2, indicating a SAW waist of [2.4 μm] at the center of the cavity, in line with the designed waist. We clearly see that as we move away from the center of the cavity, the SAW waist increases. This metric reveals a SAW waist of 4.6 μm for Dev1 [Fig. S5(d)].